\documentclass[twocolumn]{aastex631}

\begin{document}

\title{SDSO 1 is a Shocked Ghost Planetary Nebula in Front of M 31}

\author{Patrick Ogle}
\affiliation{Space Telescope Science Institute, 3700 San Martin Drive, Baltimore, MD 21218, USA}
\affiliation{Polaris Imaging}

\author{Mark Petersen}
\affiliation{4636 Terraza Mar Marvelosa, San Diego CA 92130}
\affiliation{Polaris Imaging}

\author{R. Michael Rich}
\affiliation{Dept. of Physics and Astronomy, University of California, Los Angeles, PAB 430 Portola Plaza 90095-1547}
\affiliation{Polaris Imaging}

\author{Tim Schaeffer}
\affiliation{Delft University of Technology, The Netherlands}

\author{Lewis McCallum}
\affiliation{University of St. Andrews}

\author{Alberto Noriega-Crespo}
\affiliation{Space Telescope Science Institute, 3700 San Martin Drive, Baltimore, MD 21218, USA}

\author{Biny Sebastian}
\affiliation{Space Telescope Science Institute, 3700 San Martin Drive, Baltimore, MD 21218, USA}

\author{Carl Bj{\"o}rk}
\affiliation{EPFL, Switzerland}

\author{Steeve Body}
\affiliation{Griffith University, Brisbane, Queensland, Australia}

\author{Sendhil Chinnasamy}
\affiliation{141 San Juan Dr., Georgetown, TX 78633}

\author{Marcel Drechsler}
\affiliation{\'Equipe StDr, B\"{a}renstein, Feldstra\ss e 17, 09471 B\"{a}renstein, Germany }

\author{Tarun Kottary}
\affiliation{1646 Martin Ave, Sunnyvale, CA 94087}

\author{Yann Sainty}
\affiliation{54000 Nancy, Lorraine, France}

\author{Patrick Sparkman}
\affiliation{10757 Frank Daniels Way, San Diego, CA 92131}

\author{Xavier Strottner}
\affiliation{\'Equipe StDr, Montfraze, 01370 Saint Etienne Du Bois, France}

\begin{abstract}

We present new, deep narrowband imagery and discuss the nature of SDSO 1, the large [O {\sc iii}]-emitting nebula centered $1.5\arcdeg$ SE of M 31.  We find strong evidence to support the hypothesis that SDSO 1 is unrelated to M 31 and is instead  a faded, giant ($D = 20$ pc), ghost planetary nebula (GPN) expelled by the symbiotic WD binary star EG Andromedae. The associated 45-pc long turbulent tail, seen in projection in front of M 31, yields an estimated age of 400 kyr. The initial hypersonic velocity of 91 km s$^{-1}$ drives a strong bow shock into the local interstellar medium and a reverse shock into the GPN. The SDSO 1 GPN has reached the terminal phase in its evolution where its outward expansion and forward motion have been decelerated greatly by ram pressure and the [O {\sc iii}] emission arises primarily from the reverse shock.  We establish the shock-powered GPN phase as a new phase of planetary nebula (PN) evolution, and identify 24 candidate GPNe by their large size and shock-tail morphology. This includes several giant halos of younger PNe, possibly expelled by now degenerate binary companions. The interaction of an old, fast-moving GPN with the ISM generates shocks that remain visible long after the photoionized PN shell  has faded below the limit of detectability.  

\end{abstract}


\section{Introduction} \label{sec:intro}

Planetary nebulae (PNe) give unique insights into both single and binary star stellar evolution, completing the cycle of low-mass star birth, death and return of elements to the interstellar medium (ISM).  The interaction of PNe with the ISM informs our knowledge of both the ISM and shock physics. There are several known examples of fast-moving PNe with distorted shells or bow shocks and tails indicative of PN-ISM interaction \citep{1990ApJ...360..173B, 1994AJ....108..978T, 1996ApJS..107..255T}. Sophisticated models of these systems with hydrodynamic codes  reproduce crucial features \citep{1991AJ....102.1381S, 2007MNRAS.382.1233W,  2016MNRAS.457....9C}. As we will show, PN-ISM interactions also present the opportunity to find and study otherwise undetectable old PNe by their shocks and tails. The late-stage evolution of PNe is difficult to observe because they fade rapidly as they expand \citep{2013A&A...558A..78J}. The study of of large PNe at optical wavelengths, from a few arcminutes to degrees across the sky, has a long tradition  \citep[e.g.,][]{1994AJ....108..188T, 1996ApJS..107..255T}. One of the closest examples, Sh 2-216 at 120 pc, covers as much as $1.6 \arcdeg$ in diameter \citep{1995ApJ...447..257T}.

Deep [O {\sc iii}] $\lambda 5007$~\AA~observations of M 31 and environs by amateur astrophotographers recently revealed a giant nebular arc $1.5\arcdeg$ SE of the galaxy, cataloged as SDSO 1 \citep{2023RNAAS...7....1D, 2023ApJ...957...82F}. The uncertain origins of SDSO 1 were debated, whether foreground in our own Galaxy, or directly related to M 31.  Based on its large angular size, \cite{2023ApJ...957...82F} ruled out SDSO 1 as a PN. Furthermore, they found its location above the Galactic plane and unusual structure to be inconsistent with known Galactic supernova remnants. They considered and rejected the possibility of a bow shock nebula related to a high-velocity massive star, red giant, or AGB star. Because of its large radial velocity (-95 km s$^{-1}$), EG Andromedae was considered  a likely target to develop a bow shock structure, as documented in the IRA/ISSA Survey study by \cite{1995AJ....110.2914V}. Surprisingly at the  time, the lack of warm dust was difficult to explain with the classical photoionized wind model \citep{2016AJ....152....1K}. \cite{2023ApJ...957...82F} searched for white dwarfs (WD) within $10\arcdeg$ of SDSO 1, but found none that were hot and bright enough to be a PN central star. In particular, they noted that EG Andromedae is the second-brightest UV source in the region. However, its $608 \pm 12$ pc distance (based on Gaia parallax) would imply a diameter of $20.0 \pm 0.4$ pc for SDSO 1, which they found unlikely for a PN.  

\cite{2023ApJ...957...82F} favor the explanation that SDSO 1 is an extragalactic shock associated with M 31, even though they recognize that its radial velocity of -10 km s$^{-1}$ is inconsistent with the -297 km s$^{-1}$ radial velocity of M 31. They also suggest that SDSO 1 is associated with stellar tidal streams. Large broad band surveys of M 31, like the ``Pan-Andromeda Archelogical Survey'' (PAndAS) do suggest the presence of other large emission features besides the well known stellar tidal streams \citep[][see, e.g., Fig. 1]{2009Natur.461...66M}.  However, high resolution spectroscopy yields an intrinsic [O {\sc iii}] emission line width of $\sigma<19$ km s$^{-1}$, disfavoring an extragalactic shock and instead pointing to a Galactic origin \citep{2025A&A...704A.224L}. The narrow line widths further rule out a Galactic supernova remnant, but do not rule out a shocked planetary nebula.

\cite{2025A&A...704A.224L} propose that SDSO 1 is ISM cirrus photoionized by an unknown hot star. This may be possible, but seems very unlikely to us, for the following reasons.  First, \cite{2023ApJ...957...82F} made an exhaustive search for hot stars within a 10 degree radius and found none that were hot enough, luminous enough, and nearby enough to photoionize the nebula. Any such star would have to be more than $10 \arcdeg$ away from SDSO 1 and very nearby to the sun and therefore very bright. Second, the recent Northern Sky Narrowband survey by Stefan Ziegenbalg (\url{www.simg.de/nebulae3/dr0_2}) shows little if any [O {\sc iii}] cirrus, even though SDSO 1 is clearly detected. \cite{2023ApJ...957...82F} similarly come to the conclusion that diffuse ISM is not generally a strong source of [O {\sc iii}] emission.

We present evidence that SDSO 1 may be a  very large, old, shock-powered ghost planetary nebula (GPN). We also identify 24 other known large [O {\sc iii}]-emitting nebulae and PNe giant halos as candidate GPNe. We present new narrow band imaging observations of SDSO 1 in Section 2. We detail the properties of SDSO 1 in Section 3, and its discuss its origins, dynamics, energetics, and evolution in \S4.

\section{Narrowband Imaging Observations and Data Reduction}

\begin{figure*}
  \includegraphics[trim=0.0cm 0.0cm 0.0cm 0.0cm, clip, width=0.99\linewidth]{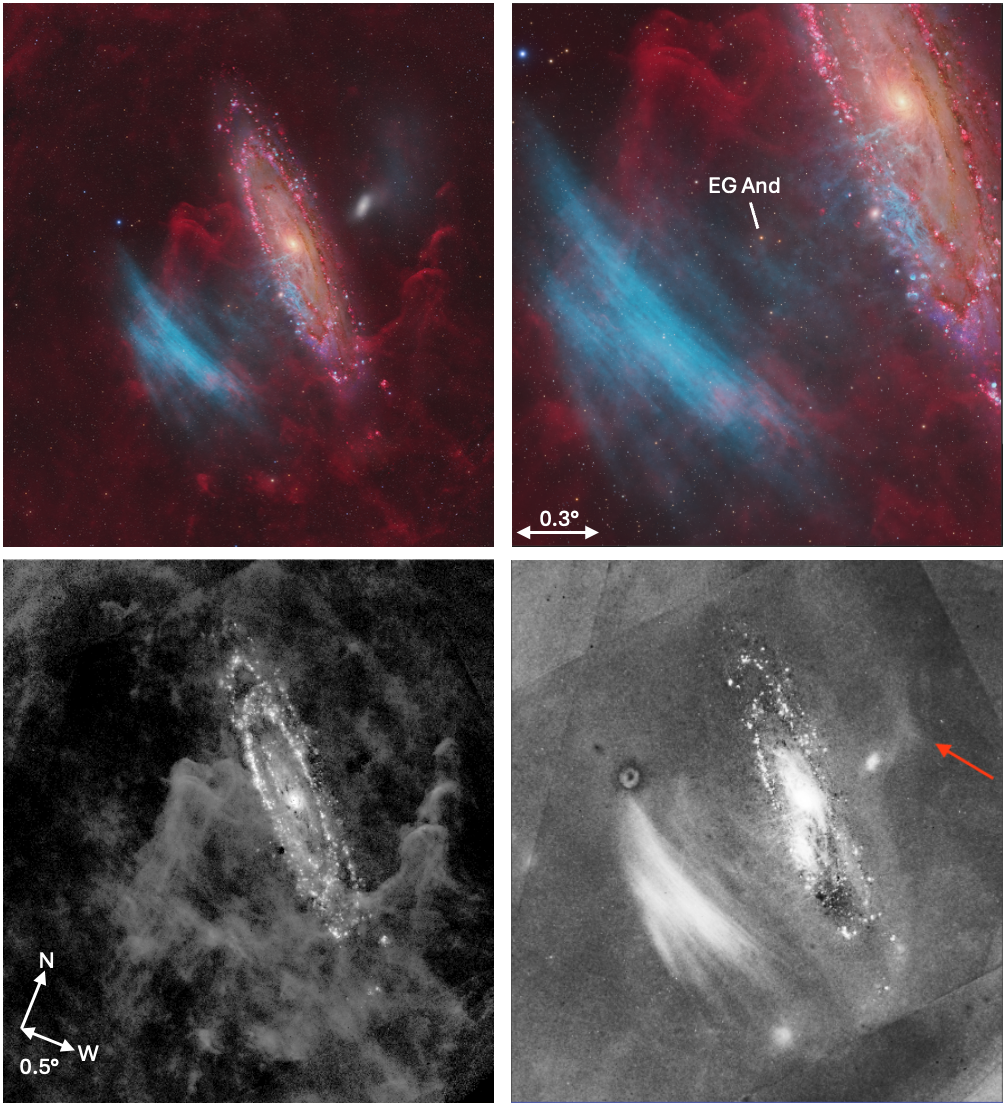}
 
\caption{Our RGB and Narrow-band imaging of SDSO 1, M 31, M 32, and NGC 205. Top Left: H$\alpha$ and  [O {\sc iii}] (HOO) plus RGB combination. Top Right: Same, zoomed in to show detail. Bottom Left: Continuum-subtracted H$\alpha$, with stars removed, using a logarithmic stretch. Bottom right: starless, continuum-subtracted [O {\sc iii}] with a hard logarithmic stretch to show the faintest features. The counter-arc location is indicated by the arrow. All images have the same orientation. }
\label{fig:M 31Arc}
\end{figure*}

\begin{figure*}
  \includegraphics[trim=0.0cm 0.0cm 0.0cm 0.0cm, clip, width=0.95\linewidth, angle = 0]{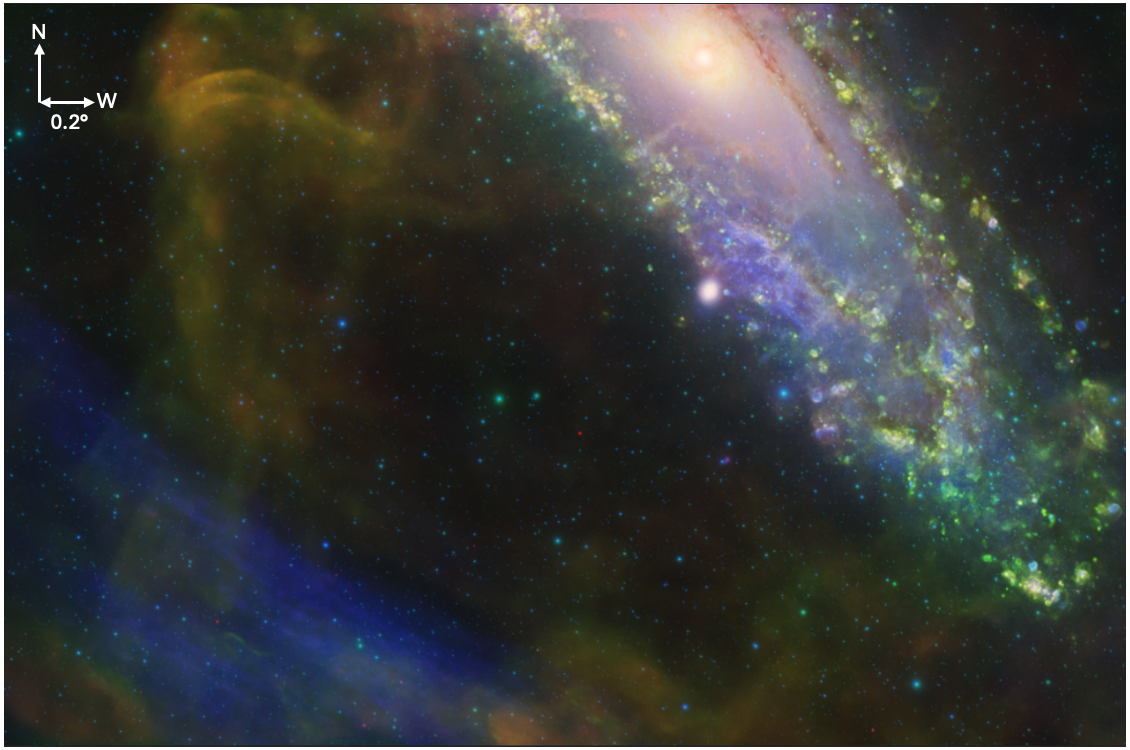}
\caption{Narrow-band (r, g, b $=$ [S {\sc ii}], H$\alpha$, [O {\sc iii}]) plus LRGB imaging of a portion of SDSO 1, M 31, and M32. Credit: Deep Sky Collective. }
\label{fig:M 31_DSC_SHO}
\end{figure*}

We observed SDSO 1 and the surrounding field of view (including M 31, M 32, and NGC 205) in narrowband [O {\sc iii}] $\lambda 5007 \AA$, H$\alpha$, and broadband R, G, and B filters (Table 1). Our deep narrowband imaging relies on long (10--30 min) exposures and CMOS sensors on high-quality refracting telescopes to accumulate enough photons to beat the readout noise. Such long exposures also require stable mounts and capable guiding systems to take advantage of the best seeing conditions. Each of five astrophotographers utilized a Takahashi FSQ 106 mm  (4-inch) refractor and focal reducer to achieve fast imaging at f/3.6 or f/3. This setup provided a wide field of view (3.5-5 degrees) covering the full extent of SDSO 1 and the disk of M 31.  The telescopes were equipped with cooled, monochrome cameras  with sensitive Sony CMOS sensors.  Observations made during the 2024 September - December season netted a total integrated exposure time of 525 hr, after culling bad frames that didn't meet our acceptance criteria.  This includes 313 hr of exposure in the [O {\sc iii}] band and 148 hr of exposure in the H$\alpha$ band. 

In order to explore additional fine details in SDSO 1, including data in the [S {\sc ii}] $\lambda 6723\AA$ band, we also present imaging over a narrower field obtained by the Deep Sky Collective for its 1000-Hour Project (released in March 2024). The equipment and integrated exposure times are listed in Table 1. This project utilized 15 telescopes with apertures ranging from 80-203 mm (3--8 inches) in diameter, obtaining 488 hr of exposure in the [O {\sc iii}] band, 314 hr in H$\alpha$, and 169 hr in [S {\sc ii}]. We utilize similar datasets for GPN candidates PN A66 15 and Hewett 1, courtesy of the NHZ collaboration and Bray Falls, respectively, so we can compare them to SDSO 1.

Basic data reduction utilized the PixInsight software suite to align, stack, and integrate the data frames. Frames demonstrating large stellar eccentricity ($>0.7$), stellar FWHM $>6$", or large median flux indicative of clouds were rejected from the integration. The Generalized Extreme Studentized Deviate (ESD) algorithm was used to reject outliers. The data frames were then optimally weighted using a custom weighting function that incorporated both S/N and stellar PSF quality metrics. Because the orientation and field of view of each wide field setup was different, extra care was taken to remove gradients using linear spline fitting functions and matching background levels at frame overlaps.

We employed broadband R observations to estimate and subtract the continuum emission at H$\alpha$ and [S {\sc ii}].  Because the wavelength dependence of galaxy stellar continuum is relatively weak in the R-band, the resulting continuum subtraction is quite robust to differences in stellar populations.  The [O {\sc iii}] band continuum subtraction on the other hand is not as clean with a single band or even a constant linear combination of the B and G bands, which flank the line. We therefore devised a new algorithm, which we call Color Continuum Subtraction (CCS), to better estimate the [O {\sc iii}] band continuum by logarithmically interpolating the continuum between the B and G bands in each pixel of our images. We dedicated extra imaging time to the B band, which is generally fainter than G against a greater sky background, in order to minimize the noise in our $ B - G$ color estimate. CCS assumes the following simple linear equation in magnitude space:

\begin{displaymath}
O_\mathrm{cont} [mag] = G_\mathrm{cont} [mag] + a + b (B-G)_\mathrm{cont}
\end{displaymath}

This log-linear color relation is calibrated against the dust-reddened stellar population colors in M 31, sampled in an aperture selected to contain both blue and red regions, but avoiding regions with nebular emission. O, G, and B image pixel values are extracted from the galaxy aperture, then the above relation is fit for its slope and intercept. The intercept contains information on the observing system relative response to photons in the 3 bands. The slope is related to the dust-reddened spectral energy distribution of the continuum and may in principal differ between different regions of a galaxy and between different galaxies. We further refined the continuum subtraction by separately fitting the CCS relation (Eqn. 1) for 3 different ranges of $B-G$, to account for small differences in intercept parameter $a$ for the youngest and oldest stellar populations.

We calibrated our [O {\sc iii}], H$\alpha$ and [S {\sc ii}] stellar photometry to {\it g} and {\it r} photometric catalog data from the Sloan Digital Sky Survey (SDSS). We then integrated the filter transmission curves to determine the filter equivalent widths for the emission line flux calibration.

\section{Results}

The deep narrowband images  reveal many details in the nebular emission from SDSO 1 (Figs. 1 \& 2). The primary [O {\sc iii}] emission extends $1.7\arcdeg$ and partially fills a circle of diameter $D_\mathrm{i} = 1.9\arcdeg$ centered on EG And. The mean surface brightness is $1.9\times 10^{-17}$ erg s$^{-1}$ cm$^{-2}$ arcsec$^{-2}$ measured within the contour at 10\% of the peak. Because the visible PN shell does not form a complete circle, we adopt this as the approximate diameter for SDSO 1.  Fainter filaments are found within a circle of diameter $D_\mathrm{o} = 2.7\arcdeg$ centered on EG And. A multitude of fine striations are present, grouped in several bundles that extend all the way to the disk of M 31, in projection. We find emission from these striations in H$\alpha$ and [S {\sc ii}] as well as [O {\sc iii}], with a strong positive correlation between the nebular emission seen in all three bands. Localized, bright H$\alpha$ features also appear to be associated with the striations seen in  [O {\sc iii}]. If associated with EG And, the [O {\sc iii}] luminosity of SDSO 1, measured within the 10\% surface brightness contour is $4.4\times 10^{33}$ erg s$^{-1}$ ($2.4 L_\odot$).

Long, undulating, H$\alpha$-emitting structures appear to form in the wake of SDSO 1, extending across and to the NW of M 31. We suggest that this H$
\alpha$ tail is physically connected to SDSO 1, providing crucial clues to its origin, motion, and evolution. These structures are often highlighted in deep H$\alpha$ images of M 31 and have in the past been attributed to non-specific, foreground H$\alpha$ cirrus in the Milky Way. The brightest two features trailing from SDSO extend up to $3.0\arcdeg$ from its leading edge ($2.0\arcdeg$ from EG And). The width of the wake, measured perpendicular to the proper motion vector of EG And, is comparable to the $1.9\arcdeg$ [O {\sc iii}] diameter of SDSO 1, providing another estimate of the transverse size of the PN shell. Additional wavy H$\alpha$ and linear dust features, may extend as far as $2.9-4.3 \arcdeg$ from the head of SDSO 1. 

We detect a new [O {\sc iii}] - emitting structure trailing SDSO 1, near NGC 205, that we dub the counter-arc,  at a distance of $1.6-1.9\arcdeg$ from EG And. This emission appears in two faint bands, separated by a band of H$\alpha$ emission. The peak surface brightness of the counter-arc is only $9.6\times 10^{-19}$ erg s$^{-1}$ cm$^{-2}$ arcsec$^{-2}$. We suggest that this is emission from the SDSO 1 tail, near the point of origin where the PN was initially ejected. Long, linear, possibly periodic H$\alpha$ bands associated with the wake may mark dynamical instabilities associated with turbulence generated by the wake. 

\section{Discussion}

\subsection{SDSO 1 as a shocked GPN} 

\begin{figure*}
  \includegraphics[trim=0.0cm 0.0cm 0.0cm 0.0cm, clip, width=0.9\linewidth]{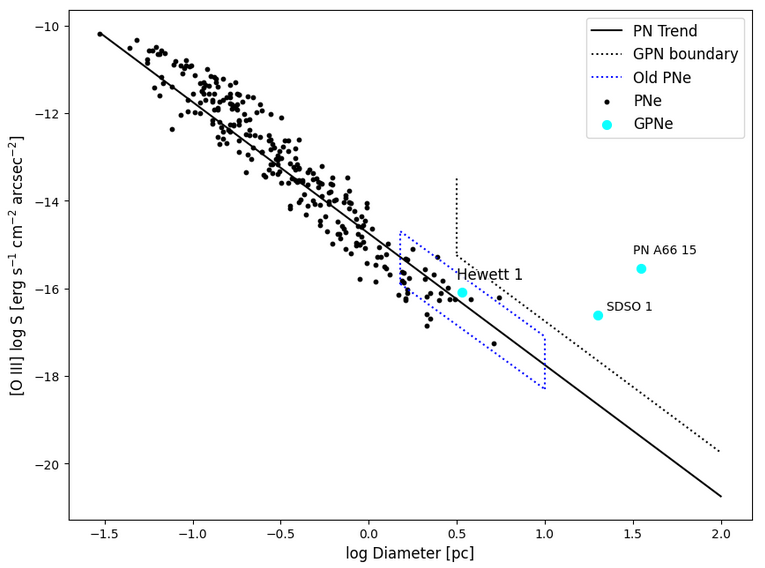}
\caption{GPNe have high [O{\sc iii}] surface brightness for their large size, compared to PNe. Mean PN surface brightness in [O {\sc iii}] is estimated from the compilation of H$\alpha$ surface brightness by \cite{2016MNRAS.455.1459F} by assuming a typical H$\alpha$/[O {\sc iii}] value of 3.3. By convention, surface brightness and diameter are measured within a contour that is $10\%$ of the peak surface brightness. PN photoionized emission surface brightness declines with diameter as $D^{-3}$ during the free expansion phase. GPNe with $>10$ times brighter [O {\sc iii}] emission above this trend are easily distinguished. Below a diameter of 3 pc, the photoionized PN shell typically dominates the emission.}
\label{fig:GPN Definition}
\end{figure*}

\begin{figure*}
  \includegraphics[trim=0.0cm 0.0cm 0.0cm 0.0cm, clip, width=0.99\linewidth]{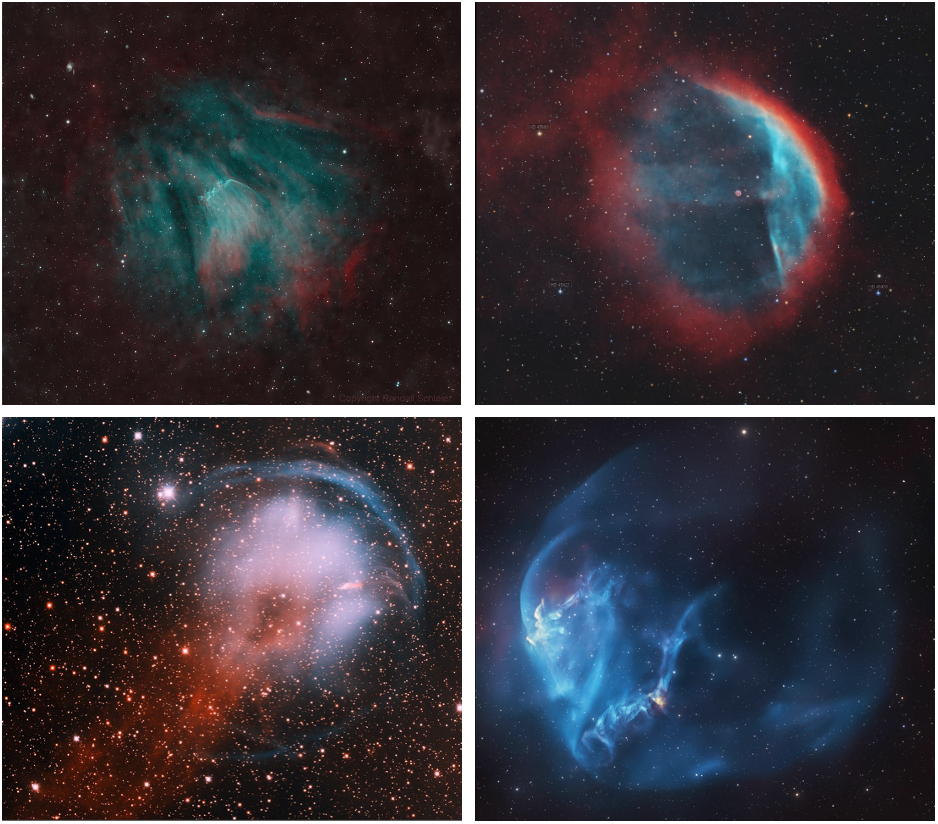}
 
\caption{GPNe and old PNe interacting with the interstellar medium. Top Left: GPN Alves 2 ($D=16$ pc; credit: Randall Schleier 2025; \url{https://app.astrobin.com/i/6rb2ww}) may be an analogue to SDSO 1. Top Right: The halo of PN A66 15 ($D = 35$ pc; credit: NHZ Team; \url{https://app.astrobin.com/i/xrz3qx}) may be an GPN + PN from a secondary WD. The strong H$\alpha$-emitting forward bow shock, [O III]-emitting reverse shock, and linear trailing features indicate fast motion through and interaction with the ISM. Bottom Left: The shock-tail structure of PN HFG 1 is also apparent, with a forward shock, contact discontinuity, and reverse shock (Credit: NOIRLab). Bottom Right: Hewett 1 \citep{2003ApJ...599L..37H} is a relatively young GPN candidate with shock-tail morphology, where the [O{\sc iii}] emission from the bow shock wraps around the remains of the mostly invisible PN, following the evolution of its expansion ($D=3.4$ pc, credit: Bray Falls, 2025; \url{https://app.astrobin.com/i/870jps}).}
\label{fig:GPNe}
\end{figure*}

 The morphology of SDSO 1, relatively high surface brightness for its size, and association with the high-velocity WD binary EG And make it likely that shocks driven by the interaction of a ghost PN (GPN) with the ISM are the primary source of the observed [O {\sc iii}] emission. As we demonstrate below, the line emission from WD photoionization is undetectable in the SDSO 1 GPN shell itself. Because they quickly expand and fade, following a $\sim t^{-3}$ density decrease \citep{1991AJ....102.1381S,2013A&A...558A..78J, 2016MNRAS.455.1459F}, observable PNe typically have diameters of $<3$ pc and ages of $<40,000$ years. However, older PNe are expected to grow to larger size and should be ubiquitous, but difficult to detect because of their low surface brightness. Together with morphological clues and central star identification, relatively high [O {\sc iii}] surface brightness is useful for distinguishing the new GPN class of PNe (Fig. 3). A GPN may identified as a nebula with diameter $D>3$ pc that is associated with a WD or hot subdwarf (HSD), and has a mean [O {\sc iii}] surface brightness more than 10 times as great as expected for its size, with respect to the trend followed by PNe. The temperature and surface brightness of shocks associated with interacting PNe are expected to depend critically on velocity with respect to the ISM \citep{1991AJ....102.1381S}. Only GPNe associated with fast-moving stars will develop shocks that are visible in [O {\sc iii}] emission.  GPNe in a narrow velocity range of 60-120 km s$^{-1}$ are expected to emit bright [O {\sc iii}] (\S 4.7) from a forward bow shock that fades once ram pressure slows their motion below this range. As the GPN decelerates, it will be overrun by a reverse shock that should also be bright in [O {\sc iii}] emission (\S 4.5). The GPN shell will eventually be disrupted and stripped from the central star, forming a long trail (\S 4.9). Fast-moving PNe in dense environments or encountering a dense cloud of gas may not survive to the GPN stage \citep{2015ApJ...799..198R}. Instead, they may be disrupted while they are still in a photoionization-dominated stage.

 Because such a small fraction of the sky had been imaged to similar depth in [O {\sc iii}], SDSO 1 was the first recognized GPN and one of only 24 GPN candidates that we have subsequently identified through a search of published images (Table 2). Its location near the famous nearby galaxy M 31 led to deep imaging of the field and its serendipitous discovery. Considering PN expansion rates and evolution \citep{2013A&A...558A..78J}, old PNe with $D = 3 - 30$ pc may be ten times more abundant than and cover 1000 times the sky area covered by $D < 3$ pc PNe. Sensitive surveys by both professional and amateur astronomers have revealed an increasing number of large, low-surface-brightness, old PNe (Table 2). The UV-bright central stars are either white dwarfs or HSDs. In many cases, the morphology consists of a nebula with a leading arc-shaped shock front, surrounded by an elongated H$\alpha$ envelope with a head-tail structure. This shock-tail morphology is a key indicator of motion through and interaction with the interstellar medium. 

Six of the GPN candidates have comparable physical size to SDSO 1, with $D>15$ pc.   Most of them have [O {\sc iii}] emission in a shock-tail morphology and are associated with fast-moving WDs (Fig. 4). Ten of the GPN candidates are giant halos of PNe that were previously thought to be diffuse ISM photoionized by EUV emission escaping the central PN \citep{2001AJ....121.1578M}. However, PN halos are typically much smaller, with $D<1$ pc \citep{1992ApJ...392..582B,2003MNRAS.340..417C}. We therefore suggest that the largest PN halos may be formed by GPNe ejected in a previous PN outflow associated with a secondary WD or HSD companion.  Discovery of GPNe via deep narrowband imaging is made possible where the emission is enhanced by ISM interaction (\S 4.5). Large GPNe that are moving slowly with respect to the ISM will remain invisible, with surface brightness much lower than the detection limit of existing [O {\sc iii}] surveys (Fig. 3).

\subsection{The SDSO 1 GPN Central Star EG Andromedae}

\begin{figure*}
 
  \includegraphics[trim=0.0cm 0.0cm 0.0cm 0.0cm, clip, width=0.95\linewidth]{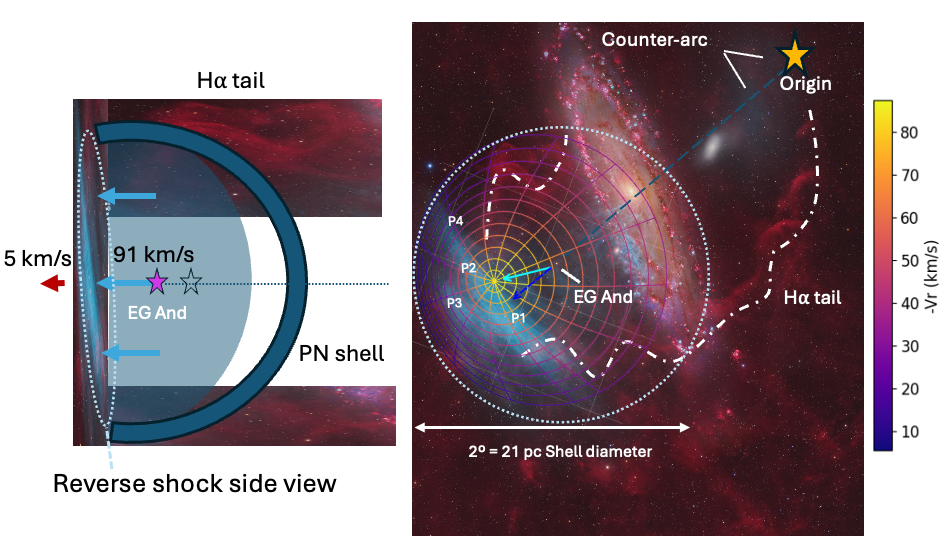}
\caption{Left: Side (deprojected) view of the compressed and shocked SDSO 1 GPN shell and H$\alpha$ tails. The reverse shock occurs where the fast-moving, low-density interior of the shell collides with the decelerated face of the shell. Compare to Figures 1 \& 2 of the simulated PN-ISM interaction by \cite{1991AJ....102.1381S}. Right: Hyperboloid geometry of the SDSO 1 GPN forward bow shock, shown for a time when it has not yet decelerated. The direction of the bowshock (cyan arrow) is opposite to the ISM headwind, which deviates from the projected proper motion of EG And. The predicted radial velocities of -35 to -85 km s$^{-1}$ are inconsistent with the values of -6 to -34 km s$^{-1}$ measured at locations P1-P4 \citep{2025A&A...704A.224L}, which are instead more consistent with the reverse shock pictured in the left panel. The H$\alpha$ tails are indicated by the dot-dashed lines.}
\label{fig:EGAnd}
\end{figure*}

The eclipsing, S-type symbiotic WD binary star EG And \citep{1950PASP...62...14W} is very likely the SDSO 1 GPN  central star, for the following reasons. EG And is the second brightest GALEX FUV source in the region, after the much lower temperature (type-B5V) star $\nu$ And.  EG And consists of a hot (50,000 - 90,000 K) white dwarf with a mass of $0.35 - 0.55 M_\odot$, orbiting a V=7.22 M2.4 III red giant with a mass of $1.1-2.4 M_\odot$  \citep{2016AJ....152....1K}. The parallax from Gaia is $1.65\pm 0.03 $ mas, indicating a distance of $608 \pm 16$ pc and a diameter of $20.0 \pm 0.4$ pc for SDSO 1.  Stellar luminosity estimates were initially made pre-Gaia, assuming a distance of $400 \pm 20$ pc, based on the properties of the red giant, including its radius measured from UV eclipses \citep{2016AJ....152....1K}. We increase the luminosity estimates by a factor of 2.31 to the Gaia distance, yielding $L_h =10 - 90 L_\odot$ for the WD and $L_g = 2300 - 4600 L_\odot$ for the red giant.  The proper motion vector of EG And ($24.8\pm 0.6$, $-44.6 \pm 1.1$) km s$^{-1}$ points towards SDSO 1, consistent with a shock   in this direction (Fig. 5).

It is important to consider the 3D location and velocity of EG And in order to understand how it can generate the SDSO 1 shock. EG And is located -22 deg (208 pc) below the plane of the Milky Way and 295 pc outside of the solar circle, at a Galactic longitude of $122\arcdeg$ (Fig. 6).  The proper motion measured by Gaia is $51 \pm 1$ km s$^{-1}$ in the plane of the sky. The  radial velocity of the M giant component of EG And ranges from -87 to  -103 km s$^{-1}$, depending on orbital phase, yielding a mean radial velocity of -95 km s$^{-1}$ \citep{2016AJ....152....1K} and a velocity vector pointing $-28\arcdeg$ along the line of sight. Accounting for the local Galactic rotation and expected lag of -5 km s$^{-1}$ above the Galactic plane, EG And lags the rotation of the Milky Way and moves hypersonically with a velocity of  -91 km s$^{-1}$ (Mach 6.1) relative to the local ISM. Its velocity, location, and age indicate that EG And is a runaway disk star with a highly elliptical orbit around the center of the Milky Way.  

The mass and age of EG And, constrained by observations of its eclipses, tell us much about the WD progenitor star and the mass it shed during its AGB and post-AGB phases. The relatively low masses of both the WD and red giant indicate low-mass progenitors, assuming that neither star was tidally stripped. However, the WD progenitor must have been slightly more massive than its M-giant companion so that it ended its time on the stellar main sequence first. Given that both stars recently (within the last 50 Myr) left the main sequence, then both the red giant and the WD progenitor must have begun as A or F dwarfs with masses of $1.1-2.4 M_\odot$ and ages of $0.5-1.5$ Gyr. The amount of mass shed by the WD progenitor would then be $\sim 0.7 - 2 M_\odot$. 

\subsection{Density and ionization of the surrounding ISM}

\begin{figure*}
  \includegraphics[trim=0.0cm 0.0cm 0.0cm 0.0cm, clip, width=0.95\linewidth]{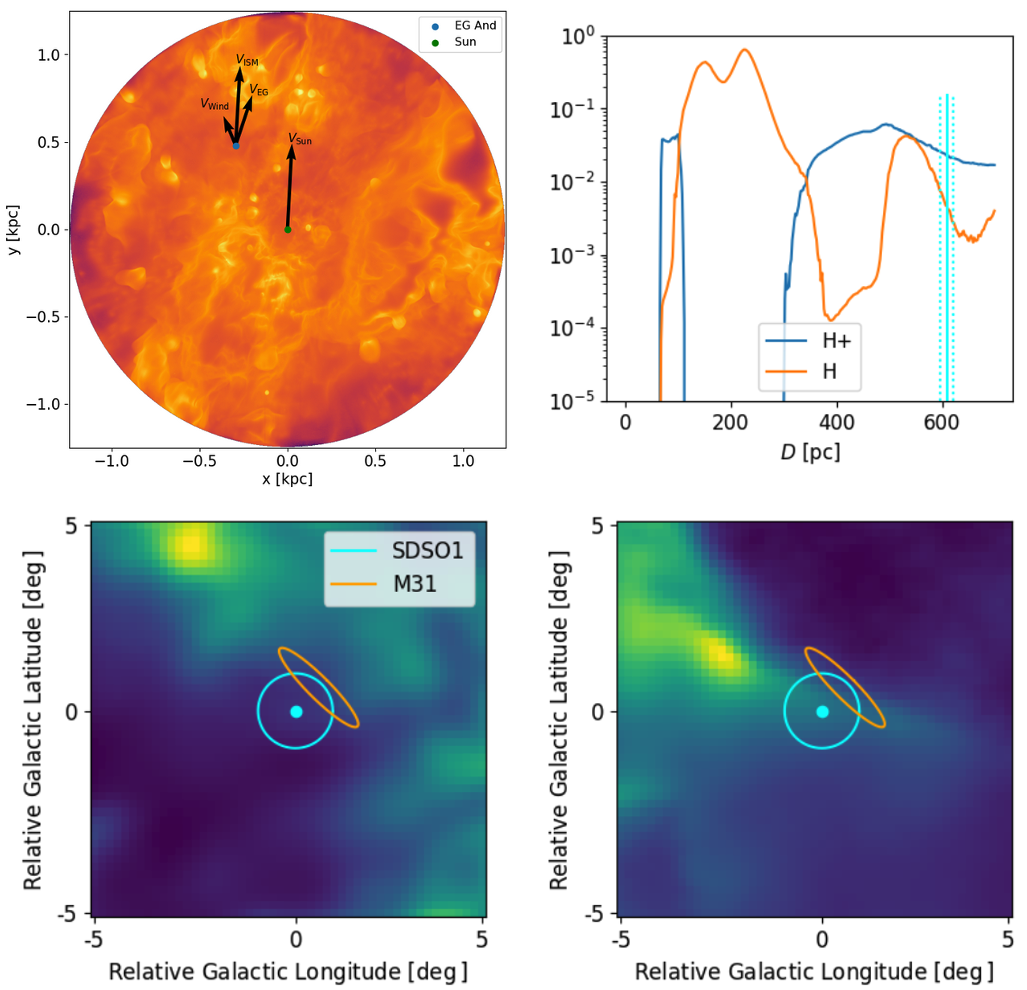}
\caption{ \cite{2025MNRAS.tmpL..22M} model of the local ISM. Top Left: Locations and velocities of EG And and the sun, plotted on a simulated map of H$\alpha$ emission, projected on the Galactic plane. EG And lags the rotation of the Milky Way, experiencing the 91 km s$^{-1}$ ISM headwind $V_\mathrm{Wind} = V_\mathrm{ISM} - V_\mathrm{EG}$. Top Right: Ionized and neutral density distributions along the line of sight in a $2\arcdeg \times 2\arcdeg$ ($40\times40$) pc$^2$ box centered on EG And. Bottom Left: Density of neutral gas in the plane of the sky. Bottom right: ionized gas density map.}
\label{fig:McCallumMap}
\end{figure*}

We use the 3D map of ionized and neutral gas by \cite{2025MNRAS.tmpL..22M} to characterize the properties of the ISM in the vicinity of EG And (Fig. 6). The map was constructed from the dust map of \cite{2024A&A...685A..82E}, and only considers ionizing photons from O stars, not B stars or WDs, so it does not consider the ionizing effects of EG And itself. EG And is located in a low-density ($n_e = 0.02$ cm$^{-3}$), ionized region above the Galactic plane, approaching a lane of neutral gas and associated ionization front which stretches across the field of view of our image. While the density gradient in its direction of motion is increasing, it is shallow enough not to significantly affect the evolution of the GPN to this point. A larger region of neutral gas seen in the map is in the foreground, at a distance of 100-300 pc from the sun. 

\subsection{The GPN Tail}

The H$\alpha$ tail of the SDSO 1 GPN is similar to those seen in fast-moving PNe such as HFG 1 \citep{1982A&A...114..414H,2009MNRAS.396.1186B,2016MNRAS.457....9C}. It likely derives from gas ablated from the PN surface by fast-moving gas at the shock-PN interface \citep{1991AJ....102.1381S,2015ApJ...805..158S}. Dividing the projected length of the SDSO 1 H$\alpha$ tail by the projected velocity of EG And yields an estimated age of 400 kyr for the SDSO 1 GPN. This estimate assumes that the full length of the GPN tail is still visible, which is supported by the apparent narrowing  and convergence of the H$\alpha$ tail at the counter-arc. It also assumes that the tail decelerates to the local ISM velocity on a timescale that is shorter than the age of the GPN. Otherwise, the length of the tail will be compressed and the age will be underestimated. Simulations of the AGB star Mira show that its  tail may decelerate slowly, leading to an underestimate of its age \citep{2007ApJ...670L.125W}. A measurement of the radial velocity of the H$\alpha$ tails would help to better constrain their age and velocity relative to the ISM surrounding EG And. 

 If SDSO 1 was expelled by EG And, it should have an initial speed of Mach 6.1 relative to the surrounding ISM, assuming a sound speed of $c_s = 15$ km s$^{-1}$ in $T=1\times 10^4$ K plasma. At that speed, the half opening angle of the Mach cone is expected to be $\theta_M = \arcsin 1/M = 9\arcdeg$. However, since the SDSO 1 GPN grew comparably as fast as it advanced, the geometry of the H$\alpha$ tail does not follow a classical Mach cone. Instead, it wraps around the back of the PN and then converges towards the counter arc. This appears to be true for the other GPN candidates too (Fig. 4). 

In addition to the H$\alpha$ tail, a number of [O {\sc iii}]-emitting filaments appear to trail behind the SDSO 1 GPN, between the location of EG And and the disk of M 31 (Fig. 1). Many of these filaments continue halfway across the face of M 31. Lacking radial velocities from spectroscopy, it is difficult to tell if all of the filaments are associated with the SDSO 1 GPN or if some if them may be ejected from the disk of M 31. We do expect to see such a tail behind the SDSO 1 GPN, similar to the tail seen in Schlieren photographs of hypersonic spheres \citep{1982VanDyke}.  A very prominent [O {\sc iii}] tail is similarly seen in GPN candidate Alves 2, and a fainter, yet very long and straight [O {\sc iii}] tail is found trailing behind the A55 15 PN (Fig. 4).

The [O {\sc iii}] emission from the feature that we call the counter-arc has never been detected before, and its origin is uncertain. Given that the counter-arc is so far removed from the head of SDSO 1, it is surprising that it shows [O {\sc iii}] emission. It could be part of the bow shock that wraps around behind the expanding GPN shell. Similar broad [O {\sc iii}] tails (shaped like backswept wings) are found in GPNe candates Alves 2, the A66 15 PN halo, and Hewett 1 (Fig. 4). Alternatively, the counter-arc may be an expansion wave, where ionized gas in the tail of SDSO 1 decelerates below the sound speed.  In that case, the alternating waves of [O {\sc iii}]  and H$\alpha$ emission might correspond to pressure waves in the ionized gas tail of SDSO 1.  

The phenomena of GPN bow shocks and tails are closely related to AGB wind bow shocks and tails \citep{1993ApJ...409..725Y,1994A&A...281L...1W,2002ApJ...572.1006Z,2006MNRAS.372L..63W,2007Natur.448..780M}. In fact, the AGB progenitors of PNe and their winds are expected to shape the ISM and set the stage for PNe-ISM interactions \citep{2007MNRAS.382.1233W}. However, because of their much lower mass outflow rate, the stand-off radii of AGB bow-shocks and their tail widths are typically much smaller than the scale of GPNe shells and tails. For example, the famous nearby AGB star Mira has a mass outflow rate of $10^{-7} M_\odot$, a bow-shock standoff radius of 0.1 pc, tail width of 0.8 pc, and observable tail length of 3.5 pc \citep{2007Natur.448..780M, 2007ApJ...670L.125W}. That means that any future Mira PN shell will overtake and merge with the AGB wind bow-shock long before transitioning to the GPN stage. While AGB wind mass outflow rates are typically much lower than PN mass outflow rates, they last a much longer time ($\sim 0.5$ Myr), and contain more mass than the PN shell. Interaction with the AGB wind may therefore considerably alter the PN shell mass, velocity, radial structure, and shape.  In our analysis below, we will assume that the combined PN plus AGB wind incorporates the entire mass shed by the star. However, it is possible that a fraction of the stellar mass was left behind in the AGB tail. It may also be possible to detect the progenitor AGB tail as a narrower extension of the GPN tail in some cases.

\subsection{GPN Expansion and Energetics}

\begin{figure*}
  \includegraphics[trim=0.0cm 0.0cm 0.0cm 0.0cm, clip, width=0.95\linewidth]{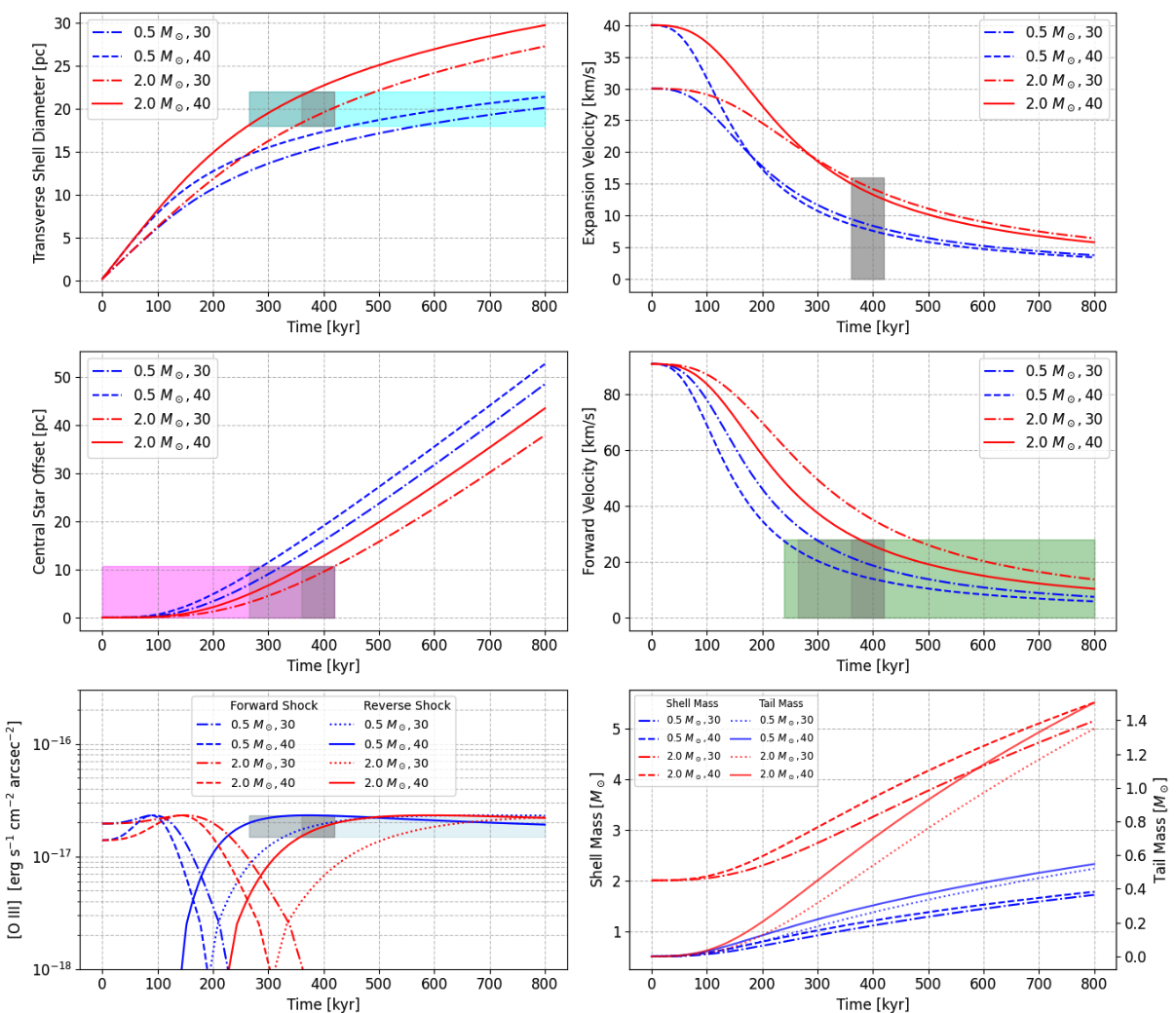}
\caption{ GPN expansion model for a shell with initial mass $0.5 - 2 M_\odot$ and expansion velocity 30-40 km s$^{-1}$, demonstrating deceleration by ram pressure. Colored bars indicate constraints on GPN age from shell diameter (cyan), central star offset (magenta), radial velocity (green), and mean [O {\sc iii}] surface brightness (pale blue). The combined constraints on age are indicated by the light grey shading, while the dark grey bar indicates our $400$ kyr age estimate from the H$\alpha$ tail length. Top left and right: Expansion slows after $\sim 100$ kyr, when the ram pressure of the ISM exceeds the ram pressure of the expanding GPN shell. Middle left: The central star subsequently becomes offset from the GPN center. Middle right: The GPN forward velocity slows to $<28$ km s$^{-1}$ after 240 kyr. Bottom left: The [O {\sc iii}] surface brightness of the forward bow shock peaks at 100-170 kyr when the [O {\sc iii}] emissivity peaks, then fades quickly as the shell decelerates. [O {\sc iii}] emission from the reverse shock predominates after about 200 kyr. Bottom right: The mass of the forward PN shell and tail grow with time as ISM is swept up.}
\label{fig:SDSO 1 expansion}
\end{figure*}

 We develop a simple model of GPN dynamics for a thin shell of mass $M$,  radius $R$, and expansion velocity $v_{exp}$ that is moving through ISM of mass density $\rho$ with velocity $v$. The expansion of the shell is slowed by ram pressure in the radial direction, while the forward motion is slowed by ram pressure  counter to the direction of motion, assuming a drag coefficient of $C_d \simeq 1.0$, appropriate for a flattened shell. The forward ram pressure increases in proportion to the cross-sectional area as the GPN expands. We assume that a constant fraction $\xi \simeq 0.7$ of the ISM that impacts the shell sticks to it, while the rest is diverted around the shell into the tail. We assume that the thermal pressure differential between the interior and exterior of the shell is negligible compared to the ram pressure. We also assume that the ongoing energy injection by EG And is negligible, based on its very weak mass outflow rate and negligible absorption of UV photons by the PN shell after its initial expulsion.  We start the clock after the PN shell has swept up and incorporated the bulk of the mass of the AGB wind.  This system is described by these ordinary differential equations:

\begin{eqnarray}
M \frac{dv_{exp}}{dt} = - \frac{1}{2} \rho v_{exp}^2 4 \pi R^2\\
M \frac{dv}{dt}= - \frac{1}{2} \rho (v + v_{exp}) ^2  C_d \pi R^2 \\
\frac{dM}{dt} =  \rho (v + v_{exp}) \xi \pi R^2\\
\frac{dx}{dt} = v
\end{eqnarray}

The GPN shell will maintain a roughly spherical shape while the forward ram pressure is negligible compared to the thermal pressure of the GPN shell. This ceases to be true when the density of the shell drops below $n_{crit} = n_0 *(v_\mathrm{star} + v_{exp})^2 /c_s^2$ and the shell begins to flatten \citep{1990ApJ...360..173B}.  For a thin PN shell of mass of $M=1.0 M_\odot$ and initial density of 10 cm$^{-3}$, this occurs early in the GPN evolution,  when $n_{PN}= n_{crit} = 1.4$ cm$^{-3}$. The forward expansion of the shell will completely stop once the density decreases to $n_{stop} = 0.2$ cm$^{-3}$, at $t_\mathrm{stop} \simeq 100$ kyr. The GPN will continue to expand laterally, flattening it even more and conforming to the shape of the forward bow-shock (Fig. 5). This justifies our assumption of a flattened shell geometry with $C_d \simeq 1.0$.

The initial velocity of the GPN is set by the 91 km s$^{-1}$ velocity of of EG And relative to the local ISM, while the ISM density of $n_e = 0.02$ cm$^{-3}$ is estimated from the \cite{2025MNRAS.tmpL..22M} model. We plot solutions for $M = 0.5 - 2.0 M_\odot$ and $v_{exp} = 30-40$ km s$^{-1}$ (Fig. 7). Higher initial expansion velocity leads to more rapid deceleration from greater ram pressure on a larger cross-sectional area. On the other hand, GPN shells with higher mass coast for longer before decelerating. Following the GPN expansion model (Fig. 7), we see that the shell expands freely for $\sim 100$ kyr, attaining a diameter of 6-8 pc. At this point, it is large enough to experience significant ram pressure and begins to decelerate both its expansion and forward motion. Together with our model, the observed SDSO 1 radial velocity of $-6$ to $-34$ km s$^{-1}$ \citep{2025A&A...704A.224L} constrains its age to $>240$ kyr,
while the observed diameter further constrains its age to $>270$ kyr. The observed displacement of EG And from the GPN center sets an upper limit on its age. While it is difficult to determine the exact center of SDSO 1, we estimate that EG And is displaced by less than one quarter of the GPN diameter, and thereby limit its age to $<420$ kyr. The combined age constraint of $270-420$ kyr is consistent with our 400 kyr age estimate from the deprojected H$\alpha$ tail length. The very large observed transverse GPN shell diameter of $20 \pm 2$ pc favors a high GPN mass ($1-2 M_\odot$). However, lower mass solutions are allowed if the initial expansion velocity is $> 40$ km s$^{-1}$.

We estimate the [O {\sc iii}] surface brightness of the forward and reverse shocks (Fig. 7) by multiplying the total kinetic power dissipated in these shocks by $\eta$, the velocity-dependent fraction of power emitted in [O {\sc iii}]:

\begin{eqnarray}
I_\mathrm{FS} = \frac{1}{2} \eta \rho (v + v_{exp})^3 /\pi \\
I_\mathrm{RS} = \frac{1}{2} \eta \rho v^3 /\pi \\
\eta = f(v)
\end{eqnarray}

The fraction of kinetic power emitted in the [O {\sc iii}] line peaks at $\eta = 0.17$ for a shock velocity of 90 km s$^{-1}$ (\S4.7). For a mass of $\sim 1.0 M_\odot$, the initial kinetic energy of the GPN relative to the ISM is $\sim 6\times 10^{46}$ erg. The mean power from ram-pressure deceleration over a 400 kyr lifetime is therefore $\sim 7\times 10^{33}$ erg s$^{-1}$, comparable to the observed [O {\sc iii}] luminosity of $2.4 L_\odot$.  This power is currently dissipated primarily by line emission from the GPN reverse shock, while [O {\sc iii}] emission from the forward shock has faded to an undetectable level (Fig. 7). The observed mean [O {\sc iii}] surface brightness is consistent with the predicted surface brightness of the reverse shock at 400 kyr for a broad range of mass and expansion velocity.

\subsection{Photoionization by EG And}

EG And is still hot and luminous enough to photoionize the SDSO 1 GPN and the surrounding ISM. While photons below the Lyman limit are subject to absorption, we find that the column density of the GPN shell is not enough to significantly impede the EUV flux. The size of the Stromgren sphere photoionized by EG And depends on its ionizing luminosity and the density of the surrounding ISM. The mean densities of ionized  and neutral gas in a $40 \times 40 $ pc box surrounding SDSO 1 are $\sim 0.02$ cm$^{-3}$ and $\sim 0.004$ cm$^{-3}$, respectively. The number of H-ionizing photons integrated over a 70,000 K blackbody with luminosity $10-90  L_\odot$ is $Q = {4\pi \over 3} r_1^3 n_H^2 \alpha_B = 1-9 \times 10^{45}$ photons, ionizing a Stromgren sphere out to distance of 15-31 pc from EG And (roughly twice the radius of the SDSO 1 GPN).  

We used Cloudy version 23.01 to model the photoionized spectrum of the SDSO 1 GPN, assuming a mass of $1 - 2 M_\odot$ is compressed into a spherical shell with thickness $\Delta R = 0.1-1.0$ pc. The actual thickness depends on the evolution of the GPN structure, including the effects of its interaction with the ISM. We assume an input black body spectrum from a WD with temperature 70 kK and luminosity $50 L_\odot$. Typical PN elemental abundances are invoked, using the Cloudy command "abundances planetary nebula" \citep{1983ApJS...51..211A}.  For these parameters, we find that the model GPN shell is only partially ionized with $N$(H+)/$N$(H$_\mathrm{tot}$) = 0.42 - 0.62, yielding negligible [O {\sc iii}] emission, and an [O {\sc iii}]/H$\beta$ ratio of 0.0041-0.062. For the specific case where $n_\mathrm{H} = n_\mathrm{stop}$ = 0.2 cm$^{-3}$, which is the expected density of the decelerated front face of the GPN shell, we find $N$(H+)/$N$(H$_\mathrm{tot}$) = 0.42 and [O {\sc iii}]/H$\beta$ = 0.023, again showing negligible photoionized [O {\sc iii}] emission. Next we modeled a filled, $2 M_\odot$ photoionized sphere, with a constant density of 0.02 cm$^{-3}$ that matches the density and pressure of the surrounding ISM. In this case, the gas is highly ionized, with $N$(H+)/$N$(H$_\mathrm{tot}$) = 0.995, yielding an [O {\sc iii}]/H$\beta$ ratio of 4.8. However, the total [O {\sc iii}] flux of $9.0\times 10^{-13}$ erg s$^{-1}$ cm$^{-2}$ is spread out over the entire projected area of the PN and the mean surface brightness of photoionized [O {\sc iii}] emission from the SDSO 1 PN ($6.1 \times 10^{-21}$ erg s$^{-1}$ cm$^{-2}$ arcsec$^{-2}$) is far below the detection limit of ground-based telescopes. This calculation emphasizes that photoionized emission the GPN is undetectable even if the central star is still capable of ionizing it. The only reason that SDSO 1 is still detectable is the strong shock driven by its interaction with the ISM. 

\subsection{Shock Models}

\begin{figure*}
  \includegraphics[trim=0.0cm 0.0cm 0.0cm 0.0cm, clip, width=0.95\linewidth]{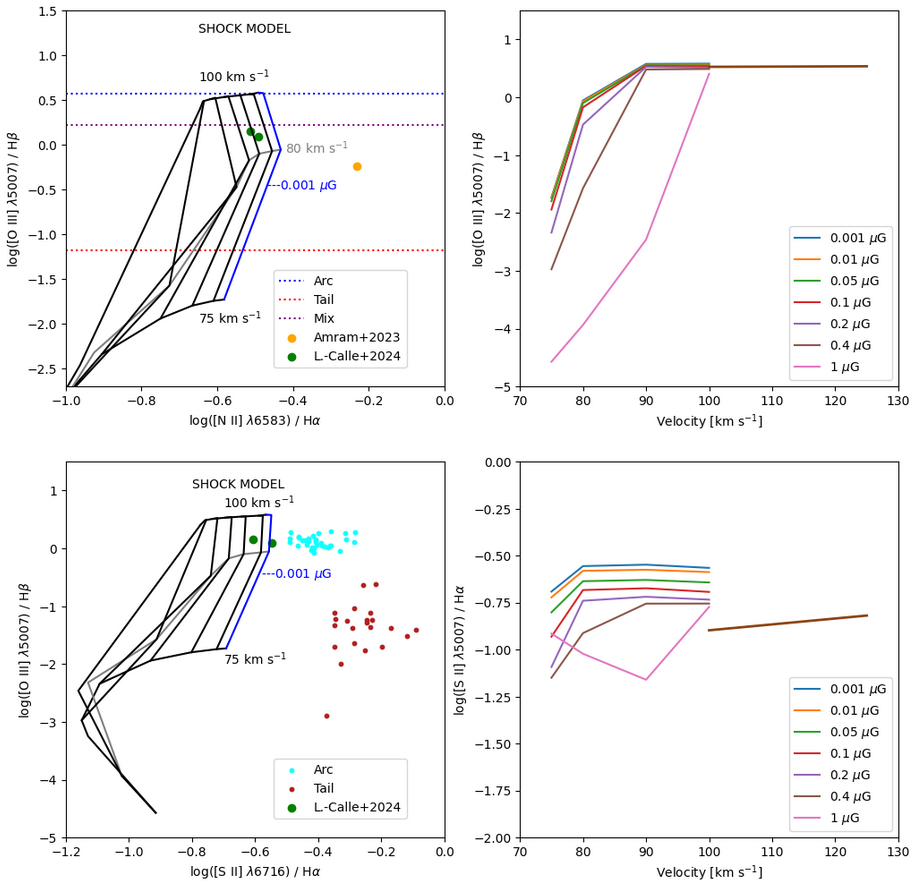}
\caption{Emission line ratios in the SDSO 1 arc and tail, compared to a grid of MAPPINGS models for a $v=75-100$ km s$^{-1}$ shock impinging on ISM with a pre-shock density of 0.01 cm$^{-3}$, solar abundance, and magnetic field of 0.001-1.0 $\mu$G. While a full velocity range of $20-200$ km s$^{-1}$ was considered, velocities below 75 km s$^{-1}$ yield little [O {\sc iii}]. Velocities in the $100-200$ km s$^{-1}$ range give similar line ratios to the 100 km s$^{-1}$ model, and were removed from the plot for clarity of presentation.  Photoionized precursor emission is not significant in this velocity range. The [O {\sc iii}]/H$\beta$ ratios for our data are estimated from the observed [O {\sc iii}]/H$\alpha$ assuming Case B recombination.  Upper and lower left: We compare our measurements and shock models to the line ratios from \cite{2025A&A...704A.224L} and \cite{2023A&A...671L..13A}. Regions where the shock is projected in front of the tail have a mixed line ratio with lower ionization as a consequence of perspective. Upper right: The [O {\sc iii}]/H$\beta$ ratio plummets for shock velocities $<80$ km s$^{-1}$. Lower right: the [S {\sc ii}]/H$\alpha$ ratio is somewhat suppressed at higher magnetic field.}

\label{fig:Shockmodel}
\end{figure*}

Our model (Fig. 7) indicates that the SDSO 1 GPN should have decelerated from its initial velocity of 91 km s$^{-1}$ (Mach 6.1) to $v < 28$ km s$^{-1}$ (Mach 1.9) at its estimated age of 400 kyr.  The corresponding projected radial velocity of $v_r < -23$ km s$^{-1}$ is consistent with the range of radial velocities $v = -6$ to -34 km s$^{-1}$ observed by \cite{2025A&A...704A.224L}, but inconsistent with the -95 km s$^{-1}$ radial velocity reported by \cite{2023A&A...671L..13A}. Because of the higher spectral resolution of the spectra of \cite{2025A&A...704A.224L}, and consistency across multiple locations in the nebula, we adopt their more accurate radial velocity measurements.

At the observed radial velocity of SDSO1, any residual forward shock should not emit perceptibly in [O {\sc iii}] (Fig. 7).  Instead, the deceleration of the front face of SDSO 1 should result in a strong reverse shock as the rest of the nebula collides with it at a relative velocity of up to 91 km s$^{-1}$ (Fig. 5). This fast reverse shock is expected to emit copious [O {\sc iii}] emission, persisting over the $\sim 2$ Myr timescale that it takes for the rear of the shell to collide with the front of the shell. Reverse shocks are also seen in younger interacting PNe as a brightening of the front edge of the PN shell \citep{1990ApJ...360..173B,1995ApJ...447..257T, 1996ApJS..107..255T, 2006MNRAS.366..387W}, and in non-interacting PNe where the fast WD wind runs into the inner edge of the PN shell \citep{2010ApJ...725.1910S}. Unlike SDSO 1, a complete shell of photoionized emission is typically still visible in such cases.

Following \cite{1991AJ....102.1381S}, the post-shock temperature for fully ionized gas is expected to be $3.45 \times 10^4 (v/50 \mathrm[km/s])^2 = 0.7-1.1 \times 10^5$ K for a reverse shock of 70-90 km s$^{-1}$ and the cooling time is $\sim 100 (v/50 \mathrm[km/s])^3/\beta n_0 $ yr $= 14-29$ kyr for an equilibrium shock with $\beta = 1.0$. We used the MAPPINGS code (ver. 5.1) to generate a grid of 1-D shock models for velocities of 20-200 km s$^{-1}$ and magnetic field $B=0.001-1$ $\mu$G \citep{2008ApJS..178...20A}.  We assume a pre-shock density of $0.01$ cm$^{-3}$, and temperature of $T = 1.0\times 10^4$ K, appropriate to the rarified interior of the GPN shell. Case B is used to convert between H$\alpha$ and H$\beta$ fluxes, assuming no reddening. We find a peak [O {\sc iii}] cooling fraction of $\eta = 0.17$, verified with Cloudy \citep{2013RMxAA..49..137F}, for a post-shock temperature of $9\times 10^4 K$ at a shock velocity of 81 km s$^{-1}$.

The emission line ratios predicted by the shock models compare favorably to the observed emission line ratios in the SDSO 1 arc  (Fig. 8). The [O {\sc iii}]/H$\beta$ emission line ratios in the highest ionization regions of SDSO 1 are consistent with the models for a  shock velocity of $80-90$ km s$^{-1}$.  The shock precursor only has a small effect on the model in this velocity range. The [S {\sc ii}]/H$\alpha$ ratio is roughly a factor of 2 higher in the SDSO 1 shock than predicted by the models. This might indicate a higher density shock or greater S abundance in the GPN than assumed in the models.  The [S {\sc ii}]/H$\alpha$ ratio is also enhanced where both the arc and tail cross the line of sight, as a consequence of projection.  The H$\alpha$ tail has much lower ionization as traced by [O {\sc iii}]/H$\beta$, possibly originating as dense gas ablated from the GPN and heated by mixing with or conduction from the shocked ISM. 

\subsection{SDSO 1 Structure}

The mildly curved, arcuate shape of the [O {\sc iii}]-brightest portion of the SDSO 1  is consistent with a GPN shell that has been compressed on its forward face, viewed only 28 degrees from the line of sight. Unlike the sharp, narrow forward shock in HFG 1 (Fig. 4), the reverse shock in SDSO 1 is rather diffuse. There is no strong [O {\sc iii}] peak, but rather a relatively uniform cap, modulated by waves of emission. Note that any finer structure might be smeared out by the relative motions of post-shock gas on the cooling timescale, corresponding to 1.2 pc (0.1 deg) in the plane of the sky. The multiple striations in SDSO 1 are reminiscent of the striations seen in other old, interacting PNe \cite{1996ApJS..107..255T}.  These striations may indicate sound waves propagating through the ionized gas or compression along magnetic field lines. Turbulence across the GPN face can  corrugate its surface and lead to vortex shedding, as in the Firefox nebula Sh 2-188 \citep{2006MNRAS.366..387W}. The wavy features in the H$\alpha$ tail of SDSO 1 look similar to this and may derive from turbulence driven by the Kelvin-Helmholtz instability. 

As discussed above, the low radial velocities observed across the face of the shock indicate a velocity similar to the expected ISM velocity of -10 km s$^{-1}$ at this location in the Galaxy \citep{2025A&A...695A.222S}. For the first half of its life, the GPN would have been capped by a fast, [O {\sc iii}]-emitting forward bow shock. The front face of the GPN shell has since decelerated to nearly the ISM velocity and emission from the bow shock has faded away. However, the down-wind nebula still follows EG And and has yet to decelerate. The fast-moving, low-density material in the GPN interior is driving a strong reverse shock as it slams into the decelerated front shell of the GPN.

In order to visualize the bow shock before it decelerated, we model it as a hyperboloid pointed in the direction of motion of EG And relative to its local ISM (Fig.~\ref{fig:EGAnd}). The asymptotic half-opening angle of the shock is set by the Mach number to be $\sim 9 \arcdeg$. There is freedom to move the head of the bow shock along its axis of symmetry, and freedom to scale the overall size of the hyperboloid to match SDSO 1. In general, the bow shock will have a nonzero angle to the velocity vector of EG And. While the proper motion of EG And is well-measured by Gaia and its radial velocity is known accurately from spectroscopic monitoring, the Galactocentric velocity of the surrounding ISM has not been accurately measured. We assume that the surrounding ISM is moving at roughly the same velocity as gas directly below it in the Galactic plane ($\sim 233$ km s$^{-1}$, but introduce an expected lag of $-5$ km s$^{-1}$ at its height of 210 pc below the plane, counter to the direction of Galactic rotation.   In this model, the projected radial velocity of the forward bow shock ranges from $-10$ to $-85$ km s$^{-1}$. The inclination of the bow shock leads to mild limb brightening at the front edge of the shock, but is not great enough for the spectroscopic lines of sight to pierce more than one side of the shock. 

The low radial velocity measurements of $-15$ to $-30$ km s$^{-1}$ by \cite{2025A&A...704A.224L}, particularly at the P1 and P2 locations, do not match the velocity of EG And and therefore cannot come from the fast forward bow shock expected for a younger ($<200$ kyr) GPN. Instead, such low-velocity [O {\sc iii}] emission can only come from a reverse shock. Since the SDSO 1 nebula is spatially resolved, the measurements by Lumbreras Calle probe the shell at distinct locations corresponding to specific velocities. In fact, there are small, significant differences of $\pm 10$ km s$^{-1}$  among different locations probed by their spectroscopy, which are consistent with the present transverse shell expansion velocity of $5-12$ km s$^{-1}$  predicted by our model.

The narrow [O {\sc iii}] line widths of $\sigma<19$ km s$^{-1}$  observed by \cite{2025A&A...704A.224L} are also consistent with a reverse shock, where the gas has decelerated from the velocity of EG And to nearly the velocity of the ISM. We established in \S4.6 that the photoionized, high velocity, pre-shock gas is undetectable in our images of SDSO 1. It would also be very difficult to detect such preshock gas and measure its radial velocity with existing spectroscopic facilities, given its predicted low surface brightness. The only detectable source of [O {\sc iiii}] emission is the post-shock gas, compressed and heated to $10^5$ K. This gas should have a velocity equal to that of the front edge of the planetary nebula, which, according to our model has decelerated to $<28$ km s$^{-1}$ with respect to the ISM. This is completely consistent with the observed velocity and line widths of $\sigma<19$ km s$^{-1}$.

While only the SDSO 1 reverse shock is currently observable in [O {\sc iiii}] and the forward bow shock appears to be invisible at its current low velocity, it should be possible to measure the radial velocity of the H$\alpha$ tails to better constrain their kinematics and history. We expect to find a  velocity gradient along the tail corresponding to the deceleration history of  SDSO1.

\subsection{Statistics and Fates of the Old PNe and GPNe Populations} 

Having identified 24 candidate GPNe so far (Table 2), we consider their importance for late-stage PN evolution. The local WD birth rate density is estimated to be $\chi_\mathrm{WD} = (1.00 \pm 0.25) \times 10^{-12}$ pc$^{-3}$ yr$^{-1}$ \citep{2005ApJS..156...47L}. Assuming that most of these WD launch PNe that are observable for $\sim 20$ kyr, and a PNe scale height of $220 \pm 25$ pc \citep{2008PhDT.......109F}, then the local PN surface density is expected to be 8.8 kpc$^{-2}$. It follows that there should be $\sim 62$ PNe associated with WD that have $D<1.5$ pc and ages $<20$ kyr within 1.5 kpc of the sun. Empirically, there are 68 PNe within 1.5 kpc with $D<1.5$ pc that have Gaia DR3 parallax measurements of their central star, from the catalog of \cite{2021A&A...656A.110C}, closely matching this expectation. Scaling the observed number by age, we expect there to be $330$ old PNe and GPNe with $D = 1.5-9$ pc and expansion ages of 17-100 kyr in this volume. However, there are only 77 in our list, compiled by cross-matching the HASH database with Gaia DR3 parallaxes (Table 2). We appear to be missing $\sim80\%$ of nearby old PNe and GPNe in this age range, which remain to be discovered, are effectively invisible because they have faded and are not interacting strongly with the ISM, or have already been stripped by an interaction with dense ISM. Depending on their survival rate, there may be as many as 1200 GPNe with expansion ages of 100-400 kyr in the same volume, but we have so far only identified 14 in this even older age range ($\sim 1\%$). 

Because their photoionized emission fades below detectability as they grow, we expect that the observability of old PNe depends in part on their velocity with respect to and interaction with the ISM. Consistent with this expectation, roughly half of old PNe (33/67) have transverse or radial velocities $>30$ km s$^{-1}$ and 24/67 have obvious interaction signatures, such as a shock-tail morphology. Furthermore, seven have measured radial velocities with $v_r>55$ km s$^{-1}$, approaching the ISM velocity required to generate copious [O {\sc iii}] emission in a shock. It will be important to measure radial velocities for more central stars to test the idea that the visibility of old PNe is enhanced by shocks driven by their motion through the ISM.

According to Fig. 7, 0.5-2 $M_\odot$ GPNe slow their expansion by 50\% by the time they reach an age of $\sim 250$ kyr. Within 800 kyr they expand to a terminal diameter of 20-30 pc. Even after its leading surface is completely flattened, the hydrodynamical simulations of \cite{1991AJ....102.1381S} demonstrate that a PN shell can potentially survive in a low density environment for $>400$ kyr in total. They estimate that the timescale for the Rayleigh-Taylor instability to disrupt the nebula if the post-shock ISM does not cool at all is about the same as the stopping time of 100 kyr. On the other hand, if the post-shock ISM cools to the same temperature as the PN shell, there is no density contrast between the two and the Rayleigh-Taylor instability is suppressed.  Furthermore, at SDSO 1's high initial Mach number of 6.1 and even at its current Mach number of ~1.5, the GPN shell is less prone to disruption by the Kelvin-Helmholtz shear instability than subsonic flow  would be \citep{2015ApJ...805..158S}. Further exploration of the complex effects and impact of the Rayleigh-Taylor and Kelvin-Helmholtz instabilities on the survivability of GPNe would require additional hydrodynamical simulations.
 
Mixing of a GPN with the ISM is eventually accomplished by ram-pressure stripping.  The reverse shock completes its propagation through the remnants of the GPN in a cloud-crushing time  of $\sim 1$ Myr, disrupting and fragmenting it. However, given the short cooling time, the remnant can persist up to 10 Myr \citep{2015ApJ...805..158S}, leaving a 100 pc-long trail as it continues to moves through the ISM at $\sim 10$ km s$^{-1}$. Given the abundance of GPNe, we predict that thousands of GPN trails may be found in sky, perhaps explaining some of the long linear features seen in all-sky H$\alpha$ surveys \citep{1998ApJ...501L..83H, 2003ApJS..149..405H}. In addition to this motion, the progenitor star will orbit the Galaxy several times and can move a large radial distance during its lifetime if given either a kick in its natal cluster or later by a companion that goes supernova. High-velocity stars can thereby serve to transport metals outward in the Galaxy disk \citep{2021MNRAS.508.4484J} and disperse them widely in the ISM via GPN outflows.

\section{Conclusions}

 Deep narrowband imaging supports a picture where SDSO 1 is a 400 kyr-old, shocked GPN. We identify the central star of the GPN as EG And, a high-velocity symbiotic WD binary at a distance of 608 pc from the sun. The GPN has expanded to a diameter of 20 pc, leading to undetectably faint WD-photoionized emission. The large diameter of the GPN led to ram-pressure deceleration, driving a $80-90$ km s$^{-1}$ reverse shock into the nebula that powers the observed [O {\sc iii}] emission. The optical emission line ratios are consistent with this reverse shock. The bright forward edge the nebula has been decelerated almost to the velocity of the ISM by the reverse shock, consistent with its observed low radial velocity and narrow emission line widths. We predict a higher radial velocity for the undetectably faint, photoionized, portion of the nebula that has yet to encounter the reverse shock. Long tails of H$\alpha$ emission trace the turbulent trail of shocked ISM left behind by the GPN. Faint filaments of [O {\sc iii}] emission found just downstream of EG And may arise in the backflow region behind the GPN. As in interacting PNe, such a shock-tail morphology seems to be prevalent in GPNe.
 
 SDSO 1 is the first recognized member of the new class of shock-powered GPNe, which are only visible by virtue of their interaction with the interstellar medium. We identify 24 candidate members of the class, including eight with diameters $>10$ pc that have likely entered a final phase, where the outflow has decelerated significantly. These GPNe are in the terminal stage of their evolution, having nearly reached their maximum diameter and will eventually be ram-pressure stripped and incorporated the surrounding interstellar medium. Several GPN candidates present as giant halos surrounding much younger PNe, suggesting that there were two PN outflows launched  100-400 kyr apart, possibly from a pair of post-AGB stars that have now become a double-degenerate WD binary.  It will be important to spectroscopically confirm or refute the presence of a binary central star in these systems, in order to test this scenario.

\begin{deluxetable*}{ll}
\tablecaption{Observations}
\tablehead{
\colhead{Band} &\colhead{Net Exposure Time (hr)} }
\startdata
\textbf{Wide-field imaging} &\\
\hline
Five telescopes: f/5 Takahashi FSQ106 EDX4 at f/3.6 or f/3 & \\
Cameras: ZWO ASI6200MM Pro, ASI2600MM Pro, ASI 6200MM; QHYCCD QHY600M, Zeus 455M Pro & \\
Narrow band Filters: Antlia \& Chroma 3 nm bandpass S, H, O    &       \\
Broad band Filters: Antlia, Chroma, \& Astrodon R, G, B & \\
$[$O \sc{iii}$]$  &  312.8\\
H$\alpha$         &  148.1\\
R                 &   18.5\\
G                 &   16.5\\
B                 &   29.5\\
Total             &  525.4\\
\hline
\textbf{Narrow-field imaging} &\\
\hline
Telescopes: APM Apo 107/700, Askar FRA600, Celestron C9.25 SC XLT, Celestron RASA 8", &\\
CFF Telescopes Refractor 135mm f/6.7, Sky-Watcher Equinox 80, Sky-Watcher Esprit 100ED, &\\
Sky-Watcher Esprit 150ED, Stellarvue SVX130T, Stellarvue SVX90T, Takahashi FSQ-106EDX4,&\\
TS-Optics Photoline 140mm f/6.5, William Optics Fluorostar 120 / FLT120, &\\
William Optics Fluorostar 132 / FLT132, William Optics ZenithStar 81 / ZS81  & \\
Cameras: QHYCCD QHY268 M, QHY600PH M; RisingCam ATR3-26000KMA; ZWO ASI2600MC Pro,&\\
ASI2600MM Pro, ASI294MM Pro, ASI6200MM Pro  & \\
Narrow band Filters: Various 3-5 nm bandpass S, H, O    &       \\
Broad band Filters: Chroma R, G, B & \\
$[$O \sc{iii}$]$ &  487.8\\
H$\alpha$         &  314.1\\
$[$S \sc{ii}$]$  &  168.9\\
R                 &   17.3\\
G                 &   26.3\\
B                 &   15.6\\
Total             & 1030.0\\
\enddata
\end{deluxetable*}

\vfil
\eject
\startlongtable
\begin{deluxetable*}{llrrlllc}
\tablecaption{Large GPNe and PNe}
\tablehead{
\colhead{Nebula} &\colhead{Central Star}  & \colhead{$v_t, v_r$$^a$} &\colhead{$d$$^b$ }& \colhead{$\Theta$$^c$} &\colhead{$D$$^c$} &\colhead{Age$^d$} &\colhead{Morph.}\\ 
\colhead{} &\colhead{}& \colhead{[km s$^{-1}$]} &\colhead{[pc]}& \colhead{[']} &\colhead{[pc]} &\colhead{[kyr]} &\colhead{} }
\startdata
{\bf GPN Candidates}\\
 NGC 7094 Halo    & WD 2134+125                    & 89,~~~-1    & 1655 &  75 & 36  & 418 & shock\\
 PN A66 15 Halo   & WD 0625-253                    &182,\nodata & 6042 &  20 & 35  & 407 & shock-tail\\
 Kn 130 Halo (OPR 1, Blue Phoenix) & GALEX J231305.2+452619& 94,\nodata & 2012 & 40 & 23  & 270 & shock\\
 SDSO 1           & EG And                         & 52,~~-95    &  608 & 113 & 20  & 232 & shock-tail\\ 
 Fal Obj 1 Halo (Kyber Crystal)& Gaia DR2 3100775558029162880 & 29,\nodata & 1502 & 40 & 18  & 202 & shock-tail\\
 Alves 2          & Gaia DR3 957591309724015872    & 210,\nodata& 1359 &  40 & 16  & 183 & shock-tail\\
 NGC 6842 Halo    & 2MASS J19550232+2917178        & 101,\nodata& 2192 &  21 & 14  & 157 & shock\\
 Pa 161 Halo      & Gaia DR3 4183333506776770688   & 69,~~~-5     & 1026 &  41 & 12  & 142 & shock\\
 MWP 1 Halo       & WD 2115+339                    & 29,\nodata &  502 &  78 & 11  & 132 & shock\\
 StDr 158 (Smaug) & V343 Serpentis                 & 69,~~~~3    & 3303 &  12 & 11  & 130 & shock-tail\\
 WPS 46 (Vulcan)  & TYC 4376-968-1                 & 10,\nodata &  971 &  38 & 11  & 125 & shock\\
 Fal 5 Halo       & Gaia DR3 5845251684417987584   & 50,\nodata& 1695 &  22 & 11  & 124 & shock\\
 PaFal 1 (God Head) & Gaia DR3 5822588589428673152 & 35,\nodata & 1582 &  23 & 11  & 123 & shock-tail\\
 NGC 3242 Halo    & GALEX J102446.3-183834         & 30,~~~5     & 1342 &  90 & 9.1 & 105 & shock-tail\\
 PaStDr 9 (Titan) & FBS 0212+385                   & 36,~~~~4     & 310  &  90 & 8.1 & 94 & shock-tail\\
 Ek 5 (Eye of Odin)& Gaia DR3 536108267549230336   &20,\nodata& 1012  & 23  & 6.8 & 78 & shock-tail\\
 EGB 10 Halo      & TYC 4454-1299-1                & 31,~~~16     &  290 &  72 & 6.1 &  70 & shock\\
 Fr 2-30          & UCAC4 686-007194               & 67, -107    &  890 &  23 & 6.0 & 69  & shock-tail\\
 Fr 2-15 (OPR 2, Zoe's Pearl)  & Gaia DR3 1755630421264083712   & 65,~~~80     &  825 &  24 & 5.8 &  67 & shock-tail\\
 StrDr Obj 33 (Azure)& Gaia DR3 2224890576564662144 &57,\nodata  &  747 &  26 & 5.7 &  65 & shock\\
 JAM 1 (MSLSOPR 1, Ringwraith)  & Gaia DR3 2041925347716149632 &36,\nodata & 1318 & 12 & 4.4 & 51 & shock\\
 Fr 2-25          & GALEX J080404.4-063058         & 23,\nodata&  824 &    17 & 4.0 & 47 & shock\\
 Fal Obj 3 (Ghast)& JL 251                         & 50,\nodata & 1053 &  11 & 3.4 &  39 & shock\\
 Hewett 1         & PG 1034+001                    & 85,~~65     &  193 &  60 & 3.4 &  39 & shock-tail\\
 \hline
{\bf Large, Old PNe ($D>1.5$ pc)}  \\
 WPS 62           & WD 0444+049                    & 53,\nodata&  440 &    60 & 7.7 & 89 &\\
 WPS 75           & GALEX J095422.9-050209         & 79,\nodata&  401 &    60 & 7.0 & 81 &\\
 Ton 320          & PG 0823+317                    & 38,~~42    &  551 &    39 & 6.3 & 73 &\\
 WPS 54            & PG 0948+534                   & 44,~~95    &  315 &    60 & 5.5 & 64 & \\
 StDr 56 (Goblet)& LAMOST J020717.41+300511.6   & 34,~~88    &  421 &    44 & 5.4 & 62 &\\
 Pa J0637.4+3327  & GALEX J063728.2+332706         &  7,\nodata& 1050 &    16 & 4.9 & 57 & shock-tail\\
 Zie 1            & PG 0038+199                    & 66,\nodata&  400 &    41 & 4.8 & 55 &\\
 EC 13290-1933    & GALEX J133146.3-194825         & 76,\nodata& 1245 &    13 & 4.7 &
 54 & \\
 Fr 2-16          & WD 2115+118                    & 11,\nodata&  520 &    30 & 4.5 & 53 &\\
 Fr 2-21          & PHL 4                          & 184,\nodata&1636&    9.0& 4.3 & 
 50 & \\
 BMP J0733-3108   & GALEX J073324.2-310804         & 28,\nodata& 1175 &    12 & 4.0 &
 46 &\\
 IsWe 2        & WD 2212+656                       & 13,\nodata&  819 &    16 & 3.8 & 44 & \\
 Sh 2-216         & WD 0439+466                    & 16,~~58    &  128 &   100 & 3.7 & 43 & shock \\
 Sh 2-176         & WD 0029+571                    & 14,\nodata& 1110 &    11 & 3.6 & 41 & shock-tail\\ 
 StDr 138      & GALEX 033755.8+434415          & 23,\nodata& 979  &    12 & 3.4 & 40 &\\
 Dr 21         & Gaia DR3 1938255660503776000   & 12,\nodata& 1543 &   7.5 & 3.4 & 39 & \\
 Sh 2-188 (Firefox)& WD 0127+581                   & 32,\nodata&  943 &    12 & 3.2 & 37 & shock-tail\\
 PFP 1            & Gaia DR3 3058094200264637312   & 24,\nodata&  534 &    19 & 3.0 &
 34 & \\
 CaVa 1        & Gaia DR2 3158200679523220096   & 22,\nodata& 1257 &    8.0& 2.9 & 34 & \\   
 WeDe 1        & WD 0556+106                    &  4,\nodata&  580 & 
 17 & 2.9 & 33 & tail\\
 BMP J1759-3321   & Gaia DR3 4042513447762858880   & 11,\nodata& 880 &   11  & 2.9 &
 33 & \\
 EGB 6            & WD 0950+139                    & 56,\nodata&  752 &   13
 & 2.8 & 33 & shock\\
 PaStDr 3      & GALEX J011928.8+490109         & 25,~~-28   &  807 &    12 & 2.8 & 33 & \\
 HDW 3 (Tiara)  & WD 0322+452                    & 30,\nodata&  916 &    10 & 2.8 & 32 & shock-tail\\
 StDr 43       & Gaia DR3 2005036900103208704   & 27,\nodata& 1301 &   7.0 & 2.6 & 31 & \\
 PN A66 74        & WD 2114+239                    & 15,\nodata& 658  & 
 13.8 & 2.6 & 31 & \\
 BMP J1808-1406   & Gaia DR3 4146962654265880064   & 11,\nodata& 1007 & 9.0  & 2.6 & 31 &\\
 Pa 161           & Gaia DR3 4183333506776770688   & 69,~~~-5   & 1026 & 8.8  & 2.6 & 30 &\\
 StDr 13       & GALEX J063422.6+072220         & 26,\nodata& 1195 & 7.5   & 2.6 & 30  & \\
 Dr 27         & GALEX J001103.6+571036         & 17,\nodata& 992  & 9.0   & 2.6 & 30  & \\
 Sh 2-290 (PN A66 31)& GALEX J085413.0+085354      & 34,\nodata&  543 &    16 & 2.6 & 30 & shock\\
 StDr 140 (Lori's)& GALEX J074320.3+160752      & 43,\nodata& 1233 &   6.9 & 2.5 &  29 & \\
 PuWe 1        & WD 0615+556                    & 41,\nodata&  395 &    21 & 2.4 & 27 &\\  
 MeWe 2-4      & GALEX J140115.5-504010         & 59,\nodata& 1159 &   7.0 & 2.4 & 27 &\\
 Dr 24         & Gaia DR2 1822392423722386482   & 35,\nodata& 1269 &   6.4 & 2.4 & 27 & tail\\
 PHR J1625-4523   & Gaia DR3 5942477172668108032   & 11,\nodata& 1475 &   5.5 & 2.4 & 27 &\\
 PN A66 29        & GALEX J084019.0-205435         & 21,\nodata& 1064 &   7.6 & 2.3 & 27 &\\
 Fal 3        & Gaia DR3 1824704383125363328    & 62,\nodata& 1658 &   4.7 & 2.3 & 26 &\\
 Kn 121       & GALEX J20420.18+135116          & 48,\nodata& 1278 &    6.0& 2.2 & 26 & shock-tail\\
 Kn 63         & Gaia DR3 3320608713128049280   & 26,\nodata& 1304 &    5.9& 2.2 & 26 & shock\\
 Sh 2-78          & WD 1900+140                    & 13,\nodata&  698 &  10.9 & 2.2 & 26 & tail\\
 Falls 1 (Eye of Ibad)& UCAC4 280-014527        & 49,\nodata&  749 &    10 & 2.2 & 25 &\\
 Alv 1         & Gaia DR3 1866922365452368768   & 33,\nodata& 1650 &    4.5& 2.2 & 25 &\\
 PN A66 21 (Medusa)& WD 0726+133                   & 26,\nodata&  592 &   12.5& 2.2 & 25 & shock-tail\\
 PN G341.5+12.1 (WR 72) & Gaia DR3 6010805807350513920 & 31,~~-93& 1225 & 6.0& 2.1 & 25 &\\
 Jacoby 1            & PG 1520+525                 & 45,\nodata&  772 &   9.5 & 2.1 & 25 &\\
 Fal 2 (Cosmic Egg)& KPD 2045+5136              & 44,\nodata& 1070 &    6.8& 2.1 & 25 & shock-tail?\\
 IPHASX J055226.2+323724& GALEX J055226.1+323724   & 27,~~11    &  981 &    7.4& 2.1 & 24 & \\
 StDr 20 (Aurore's)& Gaia DR3 3028738476748576768 & 27,\nodata&1176 &   6.0& 2.1 & 24 & shock\\
 MWP 1 (Methuselah)& WD 2115+339                   & 29,\nodata&  502 &    14 & 2.0 & 24 & shock\\
 PN A66 7         & WD 0500-15                     & 14,\nodata&  517 &    13 & 2.0 & 23 &\\
 Pa 4          & Gaia DR3 2233941236591666688   & 148,\nodata& 2191&    3.1& 2.0 & 23 &shock-tail\\
 Fr 2-23          & KPD 0311+4801                  &  7,\nodata&  231 &    28 & 1.9 & 21 &\\
 Bode 1           & GK Per                         & 39, 146   &  434 &    15 & 1.8 & 21 &\\
 FP J1824-0319    & UCAC4 434-076255               & 23,\nodata&  195 &    32 & 1.8 & 21 & shock\\
 JnEr 1 (Headphones)& WD 0753+535               &  6,\nodata&  939 &    6.6& 1.8 & 21 &\\
 Lo 1          & GALEX J025658.5-441016         & 98,\nodata&  808 &    7.5& 1.8 & 20 &\\
 StDr 99       & Gaia DR3 2210180554096762624   & 26,\nodata& 1020 &    17 & 1.8 & 20 & shock-tail\\ 
 K2-2          & Gaia DR3 3158419684195782656   & 35,~~57    &  830 &   7.3 & 1.8 & 20 &\\
 PN A66 66        & Gaia DR3 6864496115795567488   & 45,\nodata& 1148&   5.2 & 1.7 & 20 &\\
 HFG 1            & V664 Cas                       & 32,\nodata&  708 &   8.3 & 1.7 & 20 & shock-tail\\
 PaGo 2        & GALEX J174415.3+223627         & 53,\nodata&  594 &   10  & 1.7 & 20 & tail\\
 Jn 1          & WD 2333+301                    & 15,\nodata&  989 &   5.9 & 1.7 & 20 &\\
 StDr 98       & UCAC4 730-098595               & 14,\nodata&  292 &   19  & 1.6 & 19 & shock \\
 IsWe 1        & WD 0345+498                    & 40,\nodata&  425 &  12.5 & 1.5 & 18 & tail\\
 Alves 5          & GALEX J171241.6+015533         & 17,\nodata& 885 &   6.0 & 1.5 & 17 &\\
 Sh 2-174 (Valentine Rose) & GD 561                & 39,\nodata&  290 &    18 & 1.5 & 17 & shock-tail\\
\enddata
\tablenotetext{a}{Transverse and radial velocities}
\tablenotetext{b}{Parallax distance}
\tablenotetext{c}{Angular and physical diameter}
\tablenotetext{d}{Expansion age, assuming constant radial expansion at 40 km s$^{-1}$. This will tend to underestimate ages $> 100$ kyr.}
\end{deluxetable*}

\begin{acknowledgments}
Some data in this paper were obtained at Saturn Lodge observatory, adjacent to the Polaris Observatory Association.  Saturn Lodge observatory is financially supported by RMR and MP. We thank the Deep Sky Collective for providing their deep narrow-band imaging that was crucial for measuring the emission line ratios in SDSO 1. We thank the NHZ astrophotography collaboration for sharing their deep narrowband imaging of PN A66 15 and Bray Falls for sharing his deep narrowband imaging of Hewett 1.  Many thanks to Chris Wareing for reading our draft manuscript and discussions about PN-AGB wind interactions. We thank the referee for their insightful comments that led us to refine our model of GPNe. This work has made use of data from the European Space Agency (ESA) mission
{\it Gaia} (\url{https://www.cosmos.esa.int/gaia}), processed by the {\it Gaia}
Data Processing and Analysis Consortium (DPAC,
\url{https://www.cosmos.esa.int/web/gaia/dpac/consortium}). Funding for the DPAC
has been provided by national institutions participating in the {\it Gaia} Multilateral Agreement. This research has made use of the SIMBAD database, operated at CDS, Strasbourg, France.

\end{acknowledgments}

\facilities{Gaia}

\software{astropy \citep{2013A&A...558A..33A,2018AJ....156..123A},  
          Cloudy \citep{2013RMxAA..49..137F}, MAPPINGS \citep{2008ApJS..178...20A}
          }

\bibliography{ogle_sdso1}{}
\bibliographystyle{aasjournal}

\end{document}